\documentclass[graybox]{svmult}

\usepackage{mathptmx}       
\usepackage{helvet}         
\usepackage{courier}        
\usepackage{type1cm}        
\usepackage{makeidx}         
\usepackage{graphicx}        
\usepackage{multicol}        
\usepackage[bottom]{footmisc}
\usepackage{amsmath,amssymb,bbm}

\makeindex             

\begin{document}

\title*{An example of computation of the density of ones in probabilistic cellular automata by direct recursion}
\author{Henryk Fuk\'s}
\institute{Henryk Fuk\'s \at Department of Mathematics and Statistics,  Brock University, St. Catharines, Canada. \email{hfuks@brocku.ca}}
%
%
\maketitle

\abstract{
We present a method for computing probability of occurence of 1s in a configuration obtained by iteration of a probabilistic
cellular automata (PCA), starting from a random initial configuration. If the PCA is sufficiently simple, one can construct
a set of words (or blocks of symbols) which is complete, meaning that probabilities of occurence of words from this set can be expressed as linear combinations of probabilities of occurence of these words at the previous time step. One can then setup and solve  a recursion for block
probabilities.  We demonstrate an example of such PCA, which can be
viewed as  a simple model of diffusion of information or spread of rumors. Expressions for the density of ones are obtained for this rule using the proposed method.
}
\section{Introduction}

Binary probabilistic cellular automata (PCA) in one dimension are one of the most
frequently studied types of cellular automata, and one of the most natural and most frequently
encountered problems in PCA is what the author proposes to call \emph{the density response problem}:
If the proportion of ones in the initial configuration drawn from a Bernoulli distribution is $\rho_0$, what is the expected proportion of ones
after $t$ iterations of the PCA rule?

Of course, one could ask a similar question about the probability of occurence of longer blocks of symbols
after $t$ iterations of the PCA rule. Due to the complexity of PCA dynamics, it is clear that questions of this
type are rather hopeless if one wants to know the answer for an arbitrary rule. In spite of this, it may still
be possible to provide the answer if the rule is sufficiently simple. 

One of the methods which can be used to do this is studying the structure of preimages of short blocks and detecting patterns present in them.
This approach has been successfully used by the author for a number of deterministic CA rules, such as
elementary  rules 172, 142, 130 (references \cite{paper39},  \cite{paper27}, and \cite{paper40} respectively), and several others.
 It has also been
used for a special class of PCA  known as single-transition $\alpha$-asynchronous rules \cite{paper44}. 

In this chapter, however, we would like to describe yet another method of computing probabilities of blocks of 
symbols, by setting up a system of recursive equations which can then be explicitly solved. Such
a recursive system can be easily constructed for any rule for  probabilities of \emph{all} blocks, but it is normally
too big and to complex to be solved. In certain cases, however, one can find a smaller set of blocks
for which the recursion is solvable. We will present one such example, using a PCA which can be viewed 
as a simple model of diffusion of innovations or spread of rumors.

To give the reader a flavour of what to expect, let us informally define the aforementioned  PCA rule.
Suppose we have an infinite one-dimensional lattice where each site is occupied by
an individual who has already adopted the innovation (1) or who has not adopted it yet (0). Initially the
proportion of adopters is $\rho_0$.  Once the individual
adopts the innovation, he remains in state 1 forever. Individuals in state 0 change can their states to 1
(adopt the innovation) with probabilities depending on the state of nearest neighbours: if only the right (resp., left) neighbour
already adopted, the probability is $p$ (resp., $q$), and if both of them already adopted, the probability is $r$.
What is the proportion  of adopters $\rho_t$ of after $t$ iterations of the rule, assuming that the initial
configuration is drawn from a Bernoulli distribution? We will show that
the explicit formula for $\rho_t$ can derived,
\begin{equation*}
\rho_t= \begin{cases}
   1-E \left( \left( {\rho_0}\,q -1 \right)  \left( {\rho_0}\,p-1 \right) 
 \right) ^{t} -F \left( 1-r \right) ^{t} & \text{if } p q \rho_0^2-(p+q) \rho_0+r \neq 0, \\
   1 -(G+Ht)(1-r)^{t-1}      & \text{if } p q \rho_0^2-(p+q) \rho_0+r=0,
  \end{cases}
\end{equation*}
where $E,F,G,H$ are constants depending on parameters $p,q, r$ and $\rho_0$.

In order to accomplish this, we  will start from some general theoretical remarks, considering PCA as maps in the space of shift-invariant probability
measures, similarly as done in \cite{KurkaMaas2000}, \cite{Pivato2002},  \cite{Kurka2005}, and other works.
More precisely, we will look at orbits of uniform Bernoulli measures under the action of
PCA.

\section{Probabilistic cellular automata}
Probabilistic CA are often defined as stochastic dynamical systems.
In this article, we will concentrate on Boolean CA in one dimension. Let
$s_i(t)$ denote the state of the lattice site $i$ at time $t$, where
$i \in \mathbb{Z}$, $t \in \mathbb{N}$.  We will
further assume that $s_i(t) \in \{ 0, 1 \}$ and we will say that the
site $i$ is occupied (empty) at time $t$ if $s_i(t)=1$ (resp., $s_i(t)=0$).

In a probabilistic cellular automaton, lattice sites
simultaneously change states from $0$ to $1$ or from $1$ to $0$
with probabilities depending on states of local neighbours. A
common method for defining PCA is to specify a set of local
transition probabilities. For example, in order to define a
nearest-neighbour PCA one has to specify the probability
$w(s_i(t+1))| s_{i-1}(t),s_i(t),s_{i+1}(t))$ that the site
$s_i(t)$ with nearest neighbors $s_{i-1}(t),s_{i+1}(t)$ changes
its state to $s_i(t+1)$ in a single time step.

A more formal definition of nearest-neighbour PCA can be constructed as follows.
Let $r$ be  a positive integer, called \emph{radius of PCA}, and let us
consider  a set of independent Boolean random variables
$X_{i,\mathbf{b}}$, where $i \in \mathbb{Z}$ and $\mathbf{b} \in
\{0,1\}^{2r+1}$. Probability that the random variable
$X_{i,\mathbf{b}}$ takes the value $a \in\{0,1\}$ will be
assumed to be independent of $i$ and
 denoted
by $w(a|\mathbf{b})$, 
\begin{equation}
Pr(X_{i,\mathbf{b}}=a)=w(a|\mathbf{b}).
\end{equation}
Obviously, $w(1|\mathbf{b})+w(0|\mathbf{b})=1$ for all $\mathbf{b}
\in \{0,1\}^{2r+1}$. The update rule for the PCA is then defined by
\begin{equation}\label{defprobca}
  s_i(t+1)=X_{i,\{s_{i-r}(t),\ldots,s_{i}(t),\ldots,s_{i+r}(t)\}}.
\end{equation}
Note that new random variables $X$ are used at each time step $t$,
that is, random variables $X$ used at the time step $t$ are
independent of those used at previous time steps. 

Having the above definition in mind, we note that in order to fully  define a nearest-neighbour PCA rule (i.e., rule with
$r=1$), it is enough to specify eight transition probabilities $w(1|x_1x_2x_3)$ for all $x_1,x_2,x_3 \in \{0,1\}$. 
Remaining eight probabilities, $w(0|x_1x_2x_3)$, can be obtained by $w(0|x_1x_2x_3)=1-w(1|x_1x_2x_3)$.

In any dynamical system, the main object of interest is the orbit of the system starting from a given initial point,
and properties of this orbit. In the case of PCA, we often assume that the initial condition is ``random'' or
``disordered'', typically meaning that each $s_i(0)$ is set to 1 with a given probability $\rho_0$ and to $0$ 
with probability $1-\rho_o$, independently of each other. We then want to answer question of the type ``After $t$ iterations,  what is the proportion of sites in state 1?'' or ``After $t$ iterations, what is the probability of finding a pair of adjacent zeros''? In order to pose and answer questions of this kind rigorously, we will present an alternative 
definition of PCA,  as maps in the space of probability measures. 

\subsection{Orbits of probability measures}
Let ${\mathcal{A}}=\{0,1\}$ and $X={\mathcal{A}}^\mathbb{Z}$. 
A finite sequence of elements of ${\mathcal{A}}$, $\mathbf{b}=b_1b_2\ldots, b_{n}$ will be called a \emph{block}  (or \emph{word})
 of length $n$.
Set of all blocks of elements of ${\mathcal{A}}$ of all possible lengths will be denoted by ${\mathcal{A}}^{\star}$.

A \emph{cylinder set} generated by the block $\mathbf{b}=b_1b_2\ldots, b_{n}$ and anchored at $i$  is defined as
\begin{equation}
[\mathbf{b}]_i=\{ \mathbf{x}\in {\mathcal{A}}^\mathbb{Z}: \mathbf{x}_{[i,i+n)}=\mathbf{b} \}.
\end{equation}

The set of probability measures on the $\sigma$-algebra generated by cylinder sets of $X$ will be denoted by $\mathfrak{M}(X)$.
Details of construction of such measures, using Hahn-Kolmogorov theorem, can be found in \cite{paper50}. These details,
however, are not essential for our subsequent considerations. Given a probability measure $\mu \in \mathfrak{M}(X)$, measure of a cylinder set $[\mathbf{b}]_i$, denoted by
$\mu([\mathbf{b}]_i)$, is often informally called a ``probability of occurence of block $\mathbf{b}$ at site $i$''. 

Let the function $w: \mathcal{A} \times \mathcal{A}^{2r+1} \to [0,1]$, whose values are denoted by $w(a|\mathbf{b})$
for $a \in \mathcal{A}$, $\mathbf{b} \in \mathcal{A}^{2r+1}$, satisfying
$\sum_{a \in \mathcal{A}} w(a|\mathbf{b})=1$, be called \emph{local transition function}
of \emph{radius} $r$, and let its values  be called \emph{local transition probabilities}.
\emph{A probabilistic cellular automaton}  with local 
transition function $w$ is a map $F: \mathfrak{M}(X) \to \mathfrak{M}(X)$ defined as
\begin{equation} \label{rulefed}
(F\mu)([\mathbf{a}]_i)=\sum_{\mathbf{b}\in \mathcal{A}^{|\mathbf{a}|+2r}} w(\mathbf{a}| \mathbf{b}) \mu([\mathbf{b}]_{i-r})
\mathrm{\,\, for\,\, all\,\,}  i \in \mathbb{Z}, \mathbf{a} \in \mathcal{A}^{\star},
\end{equation}
where we define
\begin{equation} \label{defw}
 w(\mathbf{a}| \mathbf{b}) = \prod_{j=1}^{|\mathbf{a}|} w(a_j|b_{j}b_{j+1}\ldots b_{j+2r}).
\end{equation}
When the function $w$ takes values in the set $\{0,1\}$, the corresponding cellular automaton is called a 
\emph{deterministic CA}. 

In this paper, we will exclusively deal with shift-invariant probability measures for which $\mu([\mathbf{b}]_i)$ is independent
of $i$. We will, therefore, drop the index $i$ and simply write  $\mu([\mathbf{b}])$. Moreover, we will be interested
in orbits of Bernoulli measures $\nu_{\lambda}$ defined for $\lambda\in [0,1]$ by 
\begin{equation}
 \nu_{\lambda}([\mathbf{b}])=\lambda^{\#1(\mathbf{b})} (1-\lambda)^{\#0(\mathbf{b})} \,\,\,\, \textrm{for any} \,\,\,  \mathbf{b} \in \mathcal{A}^{\star},
\end{equation}
where $\#0(\mathbf{b})$ and $\#1(\mathbf{b})$ denote the number of zeros and ones in $\mathbf{b}$. In order to simplify the notation, we define 
\begin{equation}
P_t(\mathbf{b})=(F^t \nu_{\lambda})([\mathbf{b}]),
\end{equation}
which will be informally referred to as ``probability of occurence of block $\mathbf{b}$ after $t$ iterations
of PCA rule $F$''. With this notation, eq. (\ref{rulefed}) can be written as 
\begin{equation} \label{Prec}
 P_{t+1}(\mathbf{a})=\sum_{\mathbf{b}\in \mathcal{A}^{|\mathbf{a}|+2r}} w(\mathbf{a}| \mathbf{b}) P_t(\mathbf{b}),
\end{equation}
for any $\mathbf{a} \in \mathcal{A}^{\star}$ and $t \in \mathbb{N}$. We will furthermore define
\begin{equation} \label{P0def}
 P_0(\mathbf{a})=\nu_{\lambda}([\mathbf{a}])=\lambda^{\#1(\mathbf{a})} (1-\lambda)^{\#0(\mathbf{a})}
\end{equation}
for any $\mathbf{a} \in \mathcal{A}^{\star}$.

Elements of $\mathcal{A}^{\star}$ can be enumerated in lexicographical order, and corresponding probabilities arranged in 
an infinite column vector 
\begin{equation}
\mathbf{P}_{t}=\left(P_t(0),P_t(1), P_t(00), P_t(01),P_t(10),P_t(11),P_t(000)\ldots \right)^T.
\end{equation}
Before we continue, note that not all these probabilities are independent. Due to additivity of measure, the following 
relationships, know as consistency conditions, are valid for any $\mathbf{a}\in\mathcal{A}^{\star}$,
\begin{equation}
 P_t(\mathbf{a})=P_t(\mathbf{a}0)+P_t(\mathbf{a}1)=P_t(0\mathbf{a})+P_t(1\mathbf{a}).
\end{equation}
These conditions will be frequently used in our subsequent considerations.

Since each $P_{t+1}(\mathbf{a})$, by the virtue of eq. (\ref{Prec}), is a linear combination of a finite number of
$P_t(\mathbf{b})$ values, we can write 

\begin{equation}  \label{Precmatrix}
 \mathbf{P}_{t+1}=\mathbf{M} \mathbf{P}_t,
\end{equation}
where the infinite matrix $\mathbf{M}$ is defined by eq. (\ref{Prec}). This yields the following expression
for probabilities of all finite words, 
\begin{equation} \label{Precmatrixsol}
 \mathbf{P}_{t}=\mathbf{M}^t \mathbf{P}_0,
\end{equation}
where components of $\mathbf{P}_0$ are defined in eq. (\ref{P0def}). In theory, the above equation gives us a complete
solution of the problem of determining the orbit of Bernoulli measure under iterations of a PCA rule. In practice, however, 
computing powers of an infinite matrix is a daunting, if not impossible, task. 

In practical applications, however, we rarely need \emph{all} probabilities $P_t(\mathbf{a})$, that is, all components
of the vector   $\mathbf{P}_{t}$. Sometimes we are interested only in one specific probability, for example,
$P_t(1)$. For a binary PCA, the expected value of a given lattice site after $t$ iterations of the rule
is equal to $1 \cdot P_t(1)+ 0 \cdot P_t(0)=P_t(1)$, and for that reason, $P_t(1)$ is sometimes referred to as
 an expected  \emph{density of ones}, to be denoted by $\rho_t$,
\begin{equation}
 \rho_t=P_t(1).
\end{equation}
Note that for Bernoulli measure $\nu_{\lambda}$ we have $\rho_0=\lambda$. Given $\rho_0$, 
could one find an explicit expression for $\rho_t$ for a given PCA using eq. (\ref{Precmatrix})? This problem will be called a \textit{density response
problem}. Although it cannot be solved in a general
case, we will demonstrate that for a sufficiently simple PCA it is a doable task. 

The idea is to setup a recursion similar to $(\ref{Precmatrix})$, but using a ``smaller'' set of block probabilities,
for which the matrix $\mathbf{M}$ has somewhat simpler structure, lending itself to direct computation
of $\mathbf{M}^t$. If we could then express $\rho_t$ in terms of block probabilities from this ``smaller'' set,
we would solve the density response problem. 

Let us define the concept of the ``smaller'' set first.
A set of words $\mathcal{A}^{\star} \supset C=\{\mathbf{a}_1,\mathbf{a}_2,\mathbf{a}_3,\ldots\}$ will be called \emph{complete}
with respect to a PCA rule $F$ if for every  $\mathbf{a}\in C$ and $t\in\mathbb{N}$,  $P_{t+1}(\mathbf{a})$ can be expressed
as a linear combination of $P_t(\mathbf{a}_1),P_t(\mathbf{a}_2),P_t(\mathbf{a}_3),\ldots$.
We will show a concrete example of a complete set in the next section.

\section{Example PCA rule}
As an example, we will consider a PCA rule which generalizes some of the CA rules investigated in \cite{paper7}. This PCA can be viewed as 
a simple model for diffusion of innovations, spread of rumors, or a similar process involving transport
of information between neighbours. We consider an infinite one-dimensional lattice where each site is occupied by
an individual who has already adopted the innovation (1) or who has not adopted it yet (0). Once the individual
adopts the innovation, he remains in state 1 forever. Individuals in state 0 change can their states to 1
(adopt the innovation) with probabilities depending on the state of nearest neighbours. All changes of states
take place simultaneously. This process can be formally described as a radius 1 binary PCA with the following
transition probabilities,
\begin{align} \label{adpodef}
 w(1|000)&=0,\, w(1|001)=p,\,w(1|010)=1,\,w(1|011)=1,\\
 w(1|100)&=q,\,w(1|101)=r,\,w(1|110)=1,\,w(1|111)=1, \nonumber
\end{align}
where $p,q,r$ are fixed parameters of the model, $p,q,r \in[0,1]$. In order to illustrate the difficulty
of computing block probabilities for this rule, let us write eq. (\ref{Prec}) for blocks $\mathbf{a}$ of length 1 and 2, 
\begin{align*}
P_{t+1}(0)&=P_t(000) + ( 1-p ) P_t(001) + ( 1-q ) P_t(100) + ( 1-r ) P_t(101), \\
P_{t+1}(1)&= pP_t(001) +P_t(010) +P_t(011) +qP_t(100) +rP_t(101) +P_t(110) +P_t(111), \\
\end{align*}
\begin{align*}
P_{t+1}(00)&=P_t(0000) + ( 1-p ) P_t(0001) + ( 1-q ) P_t(1000) + ( 1-p )  ( 1-q ) P_t(1001), \\
P_{t+1}(01)&= pP_t(0001) + (1-p ) P_t(0010) + ( 1-p ) P_t(0011) +p ( 1-q ) P_t(1001)\\ &+( 1-r ) P_t(1010) + ( 1-r ) P_t(1011), \\
P_{t+1}(10)&= ( 1-q ) P_t(0100) + ( 1-r ) P_t(0101) +qP_t(1000) + ( 1-p ) qP_t(1001) \\ &+ ( 1-q ) P_t(1100) + ( 1-r ) P_t(1101), \\ 
P_{t+1}(11)&= pP_t(0010) +pP_t(0011) +qP_t(0100) +rP_t(0101) +P_t(0110) +P_t(0111) \\ &+pqP_t(1001) +rP_t(1010) +rP_t(1011) +qP_t(1100) \\ &+rP_t(1101) +P_t(1110) +P_t(1111).
\end{align*} 
As we can see, in order to know $P_{t+1}(1)$, we need to know probabilities of blocks of length 3 at time step $t$,
and in order to compute these, we would need probabilities of blocks of length 5 at time step $t-2$, etc.

We will now show, however, that for the PCA rule defined in eq. (\ref{adpodef}) a complete subset of
$\mathcal{A}^{\star}$ can be constructed. This subset consists of clusters of zeros bounded by 1 on each side, that is, of blocks
of the type $10^n1$, where $n \in \mathbb{N}$ and  $0^n$ denotes $n$ consecutive zeros.  
\begin{proposition}
The set $C=\{101,1001,100001, \ldots \}$ is complete with respect to the PCA rule defined in in eq. (\ref{adpodef}).
\end{proposition}
In order to prove this, we need to show that every $P_{t+1}(10^n1)$ can be expressed as a linear combination
of probabilities of the type $P_t(10^k1)$. Let us write 
 eq. (\ref{Prec}) for $\mathbf{a}=10^n1$. Two cases must be distinguished, $n=1$ and $n>1$. For $n=1$, we have
\begin{multline}
P_{t+1}(101)=
  p( 1-q ) P_t(01001) 
+  ( 1-r ) P_t(01010) 
+  ( 1-r ) P_t(01011)\\
+p q       P_t(10001) 
+(1-p ) q  P_t(10010)
+ ( 1-p )q P_t(10011) 
+ p( 1-q ) P_t(11001)\\
+  ( 1-r ) P_t(11010)
+  ( 1-r ) P_t(11011). 
\end{multline}
By consistency conditions, $P_t(10010)+P_t(10011)=P_t(1001)$ and $P_t(11010)+P_t(11011)=P_t(1101)$, as well
as $P_t(01001)+P_t(11001)=P_t(1001)$. This yields
\begin{multline}
P_{t+1}(101)= ( 1-r ) P_t(01010) 
+  ( 1-r ) P_t(01011)
+p q       P_t(10001)\\ 
+(1-p ) q  P_t(1001)
+ p( 1-q ) P_t(1001)
+  ( 1-r ) P_t(1101), 
\end{multline}
and further reduction is possible using $ P_t(01010) + P_t(01011) = P_t(0101)$ and 
$P_t(0101)+P_t(1101)=P_t(101)$. The final result is 
\begin{equation} \label{P101rec}
 P_{t+1}(101)=(1-r)P_t(101)+(p-2pq+q)P_t(1001)+pqP_t(10001).
\end{equation}
For $n>1$, using a similar procedure (omitted here), we obtain
\begin{equation} \label{P10n1rec}
 P_{t+1}(10^n1)=(1-p)(1-q)P_t(10^n1)+(p-2pq+q)P_t(10^{n+1}1)+pqP_t(10^{n+2}1).
\end{equation}
Equations (\ref{P101rec}) and (\ref{P10n1rec}) clearly show that the set $C$ is complete. $\square$

\section{Calculations of $P_t(10^n1)$}
Having a complete set of block probabilities, we can now write eq. (\ref{P101rec}) and (\ref{P10n1rec})
in matrix form,
\begin{equation}
\left[ \begin {array}{c} P_{t+1}(101)\\ P_{t+1}(1001) \\ \vdots \\  P_{t+1}(10^n1)
\\ \vdots \end {array} \right]
=\left[ \begin {array}{ccccccccc} 
\tilde{a} & b & c & 0 & 0 & 0 & \cdots \\
0 & a & b & c & 0 & 0 &   \\
0 & 0 & a & b & c & 0 &   \\
0 & 0 & 0 & a & b & c &   \\
\vdots &   &  &  &  &  &\ddots
 \end {array} \right]
\left[ \begin {array}{c} P_t(101)\\  P_t(1001) \\ \vdots \\  P_t(10^n1)
\\ \vdots \end {array} \right],
\end{equation}
where $a=(1-p)(1-q)$, $\tilde{a}=1-r$, $b=p-2pq+q$, and $c=pq$.

Let us define 
\begin{equation}
            \mathbf{M}=   \left[ \begin {array}{ccccccccc} 
\tilde{a} & b & c & 0 & 0 & 0 & \cdots \\
0 & a & b & c & 0 & 0 &   \\
0 & 0 & a & b & c & 0 &   \\
0 & 0 & 0 & a & b & c &   \\
\vdots &   &  &  &  &  &\ddots
 \end {array} \right] ,\,\,\,\,
\mathbf{P}_t=\left[ \begin {array}{c} P_t(101)\\  P_t(1001) \\ \vdots \\  P_t(10^n1)
\\ \vdots \end {array} \right].
\end{equation}
We will use $\mathrm{diag}(x_1,x_2,x_3,\ldots)$ to denote an infinite matrix with $x_1,x_2,x_3,\ldots$ on the diagonal and zeros elsewhere.
Similarly, $\mathrm{sdiag}(x_1,x_2,x_3,\ldots)$ will denote shifted diagonal matrix having $x_1,x_2,x_3,\ldots$ on the line above the diagonal and zeros elsewhere, and $^2\mathrm{sdiag}(x_1,x_2,x_3,\ldots)$ will denote doubly-shifted diagonal matrix, with $x_1,x_2,x_3,\ldots$ on the second line above the diagonal and zeros elsewhere. With this notation, we have
\begin{equation}
 \mathbf{M}=\mathbf{A}+\mathbf{B}+\mathbf{C},
\end{equation}
where
\begin{eqnarray}
 \mathbf{A}&=&\mathrm{diag}(\tilde{a},a,a,\ldots),\\
\mathbf{B}&=&\mathrm{sdiag}(b,b,b,\ldots),\\
\mathbf{C}&=& ^2\mathrm{sdiag}(c,c,c,\ldots).
\end{eqnarray}
Now,
\begin{equation}
 \mathbf{P}_t=\mathbf{M}^t\mathbf{P}_0,
\end{equation}
and we need to compute $\mathbf{M}^t$. We will do it by considering a special case first.
\subsection{Special case: $\tilde{a}=a$}
When $\tilde{a}=a$, matrices $\mathbf{A}$, $\mathbf{B}$, and $\mathbf{C}$ pairwise commute, thus we can use the trinomial expansion formula,
\begin{equation}
 \mathbf{M}^t=(\mathbf{A}+\mathbf{B}+\mathbf{C})^t =\sum_{i+j+k=t}\binom{t}{i,j,k}\mathbf{A}^i\mathbf{B}^j\mathbf{C}^k,
\end{equation}
where
\begin{equation}
 \binom{t}{i,j,k}=\frac{t!}{i!j!k!}.
\end{equation}
Generalizing the previously introduced notation, let
\begin{equation}
^n\mathrm{sdiag}(x_1,x_2,x_3,\ldots)
\end{equation}
denote $n$-times shifted diagonal matrix, which has $x_1,x_2,x_3,\ldots $ entries on the $n$-th line above the diagonal and zeros elsewhere.
It is straightforward to prove that
\begin{eqnarray}
 \mathbf{A}^i&=&\mathrm{diag}({a}^i,a^i,a^i,\ldots),\\
\mathbf{B}^j&=&^j\mathrm{sdiag}({b}^j,b^j,b^j,\ldots),\\
\mathbf{C}^k&=&^{2k}\mathrm{sdiag}({c}^k,c^k,c^k,\ldots),
\end{eqnarray}
and, consequently,
\begin{equation}
 \mathbf{A}^i\mathbf{B}^j\mathbf{C}^k=^{j+2k}\mathrm{sdiag}({a}^ib^jc^k,a^ib^jc^k,a^ib^jc^k,\ldots).
\end{equation}
In the first row of the above matrix, the  only non-zero element (${a}^ib^jc^k$) is in the  column $1+j+2k$.
In the second row, the  only non-zero element (${a}^ib^jc^k$) is in the column $2+j+2k$, and so on.
This means that 
\begin{equation}
 \mathbf{A}^i\mathbf{B}^j\mathbf{C}^k \mathbf{P}_0=
\left[ \begin {array}{c} 
{a}^ib^jc^k  P_0(10^{1+j+2k}1)\\ 
 {a}^ib^jc^k P_0(10^{2+j+2k}1) \\ 
 {a}^ib^jc^k P_0(10^{3+j+2k}1) \\ 
 \vdots \end {array} \right].
\end{equation}
Using the above and the fact that $P_0(10^n1)={\rho_0^2}(1-{\rho_0})^n$, we can now write
\begin{equation}
 \mathbf{P}_t=\mathbf{M}^t\mathbf{P}_0=\sum_{i+j+k=t} \binom{t}{i,j,k}
\left[ \begin {array}{c} 
\tilde{a}^ib^jc^k  {\rho_0^2}(1-{\rho_0})^{1+j+2k}\\ 
 {a}^ib^jc^k {\rho_0^2}(1-{\rho_0})^{2+j+2k} \\ 
 {a}^ib^jc^k {\rho_0^2}(1-{\rho_0})^{3+j+2k} \\ 
 \vdots \end {array} \right].
\end{equation}
We finally obtain
\begin{multline} \label{P10l1special}
 P_t(10^l1)=\sum_{i+j+k=t} \binom{t}{i,j,k} {a}^ib^jc^k  {\rho_0^2}(1-{\rho_0})^{l+j+2k}\\
={\rho_0^2}(1-{\rho_0})^l\sum_{i+j+k=t} \binom{t}{i,j,k} {a}^i[b(1-{\rho_0})]^j[c(1-{\rho_0})^2]^k\\
={\rho_0^2}(1-{\rho_0})^l\left({a}+b(1-{\rho_0})+c(1-{\rho_0})^2\right)^t.
\end{multline}

\subsection{General case}
We are now ready to handle the general case, without the $\tilde{a}=a$ assumption. Let us first note that $t$-th powers of matrices 
\begin{equation}
\left[ \begin {array}{ccccccccc} 
\tilde{a} & b & c & 0 & 0 & 0 & \cdots \\
0 & a & b & c & 0 & 0 &   \\
0 & 0 & a & b & c & 0 &   \\
0 & 0 & 0 & a & b & c &   \\
\vdots &   &  &  &  &  &\ddots
 \end {array} \right]^t, \,\,\,\,  
\left[ \begin {array}{ccccccccc} 
a & b & c & 0 & 0 & 0 & \cdots \\
0 & a & b & c & 0 & 0 &   \\
0 & 0 & a & b & c & 0 &   \\
0 & 0 & 0 & a & b & c &   \\
\vdots &   &  &  &  &  &\ddots
 \end {array} \right]^t 
\end{equation}
differ only in their first row. This implies that the expression for $P_t(10^l1)$ given in eq. (\ref{P10l1special}) remains valid 
for $l>1$ even if $\tilde{a}\neq a$. We only need to consider $l=1$ case, that is, to compute $P_t(101)$. This can be done by
writing eq. (\ref{P101rec}) and replacing $P_t(1001)$ and $P_t(10001)$ by appropriate expressions obtained from eq. (\ref{P10l1special}),
\begin{multline} \
 P_{t+1}(101)=\tilde{a} P_t(101)
+b{\rho_0^2}(1-{\rho_0})^2\left({a}+b(1-{\rho_0})+c(1-{\rho_0})^2\right)^t\\
+c {\rho_0^2}(1-{\rho_0})^3\left({a}+b(1-{\rho_0})+c(1-{\rho_0})^2\right)^t.
\end{multline}
This can be written as
\begin{equation}  \label{nonhomo}
 P_{t+1}(101)=\tilde{a} P_t(101) + K \theta^t,
\end{equation}
where 
\begin{align}
K&=b{\rho_0^2}(1-{\rho_0})^2 +c {\rho_0^2}(1-{\rho_0})^3,\\
\theta&={a}+b(1-{\rho_0})+c(1-{\rho_0})^2.
\end{align}
Eq. (\ref{nonhomo}) is  a first-order non-homogeneous difference equation for $P_t(101)$, and, as such, it can be easily solved by standard methods \cite{Cull2004}. The solution is
\begin{equation}
 P_t(101)= P_0(101) \tilde{a}^t + K  \sum_{i=1}^t \tilde{a}^{t-i} \theta^{i-1}.
\end{equation}
The sum on the right hand side is a partial sum of geometric series if $\tilde{a}\neq \theta$, 
or of an arithmetic series when  $\tilde{a} = \theta$. Using appropriate formulae for partial sums of geometric
and arithmetic series one obtains
\begin{equation}
 P_t(101) = \begin{cases} 
 P_0(101) \tilde{a}^t  +K  (\tilde{a}^t-\theta^{t})/(\tilde{a}-\theta)    & \text{if }	\tilde{a} \neq \theta,\\
 P_0(101) \tilde{a}^t + K \tilde{a}^{t-1}t & \text{if }	\tilde{a} = \theta.
 \end{cases}
\end{equation}
Taking $P_0(101)=\rho^2(1-\rho)$ and replacing $K$ and $\theta$ by their definitions we obtain,  for $a-\tilde{a}+b (1-{\rho_0})+c (1-{\rho_0})^2 \neq 0$ (which is equivalent to $\tilde{a}\neq\theta$), 
\begin{multline} \label{P101sola}
 P_t(101)=
\frac {{{\rho_0}}^{2} \left( 1-{\rho_0} \right)^2 \left( b+c-c{\rho_0} \right)   }
{a-\tilde{a}+b (1-{\rho_0})+c (1-{\rho_0})^2}
\left( a+b (1-{\rho_0})+c (1-{\rho_0})^2
 \right)^{t}  \\
+\frac {{{\rho_0}}^{2} \left( 1-{\rho_0} \right)  \left( a-\tilde{a} \right) }
{a-\tilde{a}+b (1-{\rho_0})+c (1-{\rho_0})^2}
\, {\tilde{a}}^{t}.
\end{multline}
For $a-\tilde{a}+b (1-{\rho_0})+c (1-{\rho_0})^2=0$  (equivalent to $\tilde{a}=\theta$), the solution is slightly simpler,
\begin{equation} \label{P101solb}
 P_t(101)=\rho_0^{2} \left( 1-\rho_0 \right)  \left(  \left( c{\rho_0}
^{2}- \left( b+2\,c \right) \rho_0+b+c \right) t+\tilde{a} \right) {\tilde{a}}^{t-1}.
\end{equation}

We now have expressions for $P_t(10^l1)$ for $l=1$ (eq. \ref{P101sola} and \ref{P101solb}) and for  $l>1$ (eq.  \ref{P10l1special}).

\section{Cluster expansion}
We are finally ready to compute  $\rho_t$. To do this, we will use the formula
\begin{equation} \label{clusterexpansion}
 P_t(0)=\sum_{k=1}^\infty k P_t(10^k1),
\end{equation}
which we will refer to as ``cluster expansion''.
Various proofs of this formula can be given (see, for example, \cite{Stauffer94}), but we will show here that
it is a direct consequence of additivity of measure.

Consider a cylinder set of a single zero anchored at $i$, $[0]_i$. A single zero must belong to a cluster of zeros of size $k$ with possible values of  $k$ varying from
$1$ to infinity. If it belongs to a cluster of $k$ zeros, than it must be the $j$-th zero of the cluster, with  possible values of $j$ varying from
1 to $k$. Therefore,
\begin{equation}
 [0]_i=\bigcup_{k=1}^{\infty} \bigcup_{j=1}^{k} [10^k1]_{i-j}.
\end{equation}
Since all the cylinder sets on the right hand side are mutually disjoint, their measures add up, thus
\begin{equation}
 (F^t\nu_{\lambda})([0]_i)=\sum_{k=1}^{\infty} \sum_{j=1}^{k}  (F^t\nu_{\lambda}) ([10^k1]_{i-j}).
\end{equation}
The measure is shift-invariant, thus $(F^t\nu_{\lambda}) ([10^k1]_{i-j})=P_t(10^k1)$, and we obtain
\begin{equation}
 P_t(0)=\sum_{k=1}^{\infty} \sum_{j=1}^{k}  P_t(10^k1),
\end{equation}
which yields eq. (\ref{clusterexpansion}), as desired.

 We can now compute $P_t(0)$ using the cluster expansion formula and eq. (\ref{P101sola}),  (\ref{P101solb}) and  (\ref{P10l1special}).
We will first consider the case of  $a-\tilde{a}+b (1-{\rho_0})+c (1-{\rho_0})^2 \neq 0$, that is, using  eq. (\ref{P101sola})
for $P_t(101)$.
 \begin{multline}
  P_t(0)=\sum_{l=1}^{\infty} lP_t(10^l1)
 =
\frac {{{\rho_0}}^{2} \left( 1-{\rho_0} \right)^2 \left( b+c-c{\rho_0} \right)   }
{a-\tilde{a}+b (1-{\rho_0})+c (1-{\rho_0})^2}
\left( a+b (1-{\rho_0})+c (1-{\rho_0})^2
 \right)^{t}  \\
+\frac {{{\rho_0}}^{2} \left( 1-{\rho_0} \right)  \left( a-\tilde{a} \right) }
{a-\tilde{a}+b (1-{\rho_0})+c (1-{\rho_0})^2}
\, {\tilde{a}}^{t}
 + \sum_{l=2}^{\infty} l {\rho_0^2}(1-{\rho_0})^l\left(a+b(1-{\rho_0})+c(1-{\rho_0})^2\right)^t.
 \end{multline}
 Since 
 \begin{equation}
  \sum_{l=2}^{\infty} l (1-{\rho_0})^l=\frac {(1+{\rho_0})(1-{\rho_0})^2}{{{\rho_0}}^{2}},
 \end{equation}
 we obtain
\begin{multline}\label{nondegenerateP0}
  P_t(0)=
\frac {{{\rho_0}}^{2} \left( 1-{\rho_0} \right)^2 \left( b+c-c{\rho_0} \right)   }
{a-\tilde{a}+b (1-{\rho_0})+c (1-{\rho_0})^2}
\left( a+b (1-{\rho_0})+c (1-{\rho_0})^2
 \right)^{t}  \\
+\frac {{{\rho_0}}^{2} \left( 1-{\rho_0} \right)  \left( a-\tilde{a} \right) }
{a-\tilde{a}+b (1-{\rho_0})+c (1-{\rho_0})^2}
\, {\tilde{a}}^{t}
 + (1+{\rho_0})(1-{\rho_0})^2   \left(a+b(1-{\rho_0})+c(1-{\rho_0})^2\right)^t.
 \end{multline}
After substitution of $\tilde{a}, a, b, c$ and simplification, as well as taking $\rho_t=1-P_t(0)$,  the following 
expression for $\rho_t$ is obtained,
\begin{multline} \label{degenerateP0}
\rho_t=1-
\frac {\left( 1-{\rho_0} \right)^2  \left( r- \left( p-r+q \right) {\rho_0}\right)  }
{pq{{\rho_0}}^{2}- \left( p+q
 \right) {\rho_0}+r}   \left( \left( {\rho_0}\,q -1 \right)  \left( {\rho_0}\,p-1 \right) 
 \right) ^{t} \\
-\frac { {{\rho_0}}^{2} \left( 1-{\rho_0} \right)  \left(  \left( q-1 \right) p-q+r \right) 
 }
{pq{{\rho_0}}^{2
}- \left( p+q \right) {\rho_0}+r} \left( 1-r \right) ^{t}.
\end{multline}

When $a-\tilde{a}+b (1-{\rho_0})+c (1-{\rho_0})^2 =0$, similar calculations can be performed, but this time  using
using  eq. (\ref{P101solb}) for $P_t(101)$. After simplification, this yields
\begin{equation}
\rho_t= 1- \left( 1-\rho_0 \right) \left( \rho_0^{2} \left(\rho_0 -1\right) 
 \left( p q \rho_0-p+p q-q \right) t+1-r \right)   \left( 1-r \right) ^{t-1}.
\end{equation}
Let us summarize this in a more readable form, noticing that after substitution of $\tilde{a},a,c,b$ by their definitions the condition $a-\tilde{a}+b (1-{\rho_0})+c (1-{\rho_0})^2 =0$
becomes $p q \rho_0^2-(p+q) \rho_0+r=0$. Our final expression for the density of ones can be written as
\begin{equation}
\rho_t= \begin{cases}
   1-E \left( \left( {\rho_0}\,q -1 \right)  \left( {\rho_0}\,p-1 \right) 
 \right) ^{t} -F \left( 1-r \right) ^{t} & \text{if } p q \rho_0^2-(p+q) \rho_0+r \neq 0, \\
   1 -(G+Ht)(1-r)^{t-1}      & \text{if } p q \rho_0^2-(p+q) \rho_0+r=0,
  \end{cases}
\end{equation}
where definitions of $E,F,G,H$ can be figured out by comparing the above to eq. (\ref{nondegenerateP0}) and (\ref{degenerateP0}).

We can see that in the non-degenerate case (when $p q \rho_0^2-(p+q) \rho_0+r \neq 0$), the limit $\rho_{\infty}=\lim_{t\to \infty}\rho_t$ always exists, and that $\rho_t$ approaches 
$\rho_{\infty}$ exponentially fast, excluding special cases when $\rho_t=\mathrm{const}$ (such as
$\rho_0=0$ or $p=q=0$, $r=1$). In the degenerate case,  $\rho_{\infty}$ always exists as well, but the approach of $\rho_t$ to $\rho_{\infty}$
is linearly-exponential.

It is worth noting that the existence of the degenerate case is a fairly subtle phenomenon, and that it would be very difficult to
discover the linearly-exponential convergence  by computer simulations alone. This illustrates the point that having a  formula for $\rho_t$ brings
some advantages, and that the search for such formulae is worthwhile.

As a separate remark, let us note that deterministic CA are nothing else but special cases of PCA, thus we can choose integer values of 
$p,q,r$ and obtain relevant expression for $\rho_t$ for a number of elementary CA rules (ECA), as follows.\\
ECA rule 206 ($p=1,q=0,r=0$) or rule 220 ($p=0,q=1,r=0$)
\begin{equation}
\rho_t= 1-\rho_0(1-\rho_0)-(1-\rho_0)^{t+2},
\end{equation}
ECA rule 222 ($p=q=1,r=0$)
\begin{equation}
\rho_t=1+\frac{2(1-\rho_0)^{2t+1} + \rho_0 (1-\rho_0)}{\rho_0-2},
\end{equation}
ECA rule 236 ($p=0,q=0,r=1$)
\begin{equation}
\rho_t= 1-(\rho_0+1)(1-\rho_0)^2,
\end{equation}
ECA rule 238 ($p=1,q=0,r=1$) or rule 252 ($p=0,q=1,r=1$) ,
\begin{equation}
\rho_t= 1-(1-\rho_0)^{t+1},
\end{equation}
ECA rule 254 ($p=q=r=1$)
\begin{equation}
\rho_t= 1-(1-\rho_0)^{2t+1}.
\end{equation}
The above formulae agree with those derived informally in \cite{paper7}.
\section{Conclusions}
We presented a method for computing the density of ones in the orbit of the Bernoulli measure under the action of
a probabilistic cellular automaton, using a simple PCA rule as an example. For this rule, we were able to construct 
a complete set of block probabilities, and then solve the resulting recurrence relations. By using the cluster expansion, we then obtained the 
required density of ones. 
Although this method is obviously applicable only to PCA rules with rather simple dynamics, it may be possible to find other
PCA rules with complete sets, thus making the method useful for them. Generalization of the rule used in this paper to larger 
neighbourhood sizes comes to mind as a first possibility, and sufficiently simple deterministic rules, such as asymptotic emulators of identity investigated in 
\cite{paper52}, are also possible candidates. 

\section{Acknowledgments}
The author acknowledges partial financial support from the Natural
Sciences and Engineering Research Council of Canada (NSERC) in the
form of Discovery Grant. Some calculations on which this work is based were made
 possible by the facilities of the Shared
Hierarchical Academic Research Computing Network (SHARCNET:www.sharcnet.ca) and
Compute/Calcul Canada. The author thanks anonymous referees for suggestions leading to improvement of the article,
including a simpler derivation of the cluster expansion formula.

\end{document}